\documentclass{article}
\usepackage{lajolla2006}
\usepackage{graphicx}
\usepackage{subfig}

\frompage{000} \topage{000}                                              
\def\rightmark{Electrons and Heavy Quark at PHENIX Detector}
\def\leftmark{C. L. Silva for the PHENIX Collaboration}
\def\pp{$p$+$p~$}
 
\title{Electrons and Heavy Quark at PHENIX Detector.} 
\authors{
{Cesar Luiz da Silva$^{1,a}$ for the PHENIX Collaboration %
}\\[2.812mm]
{\normalsize
\hspace*{-8pt}$^1$ University of S\~ao Paulo, \\ 
S\~ao Paulo, Brazil\\[0.2ex] 
}}
 
\abstract{Measurement of heavy quark production is one of the tools used
to investigate the  matter produced in extremely hot and dense
conditions in heavy ion collisions at RHIC.
The PHENIX experiment has measured mid-rapidity transverse
momentum spectra of electrons.  After subtracting the photonic
background contribution, the electron spectra are mainly due to semileptonic
decays of hadrons containing heavy quarks and therefore provide a
measurement of  heavy quark production and its energy loss
in hot and dense matter. This paper will present the technique used by the PHENIX experiment
and recent results on heavy quark production in \pp, d+Au and
Au+Au  collisions at $\sqrt{s_{NN}} = $200 GeV.}

 
\begin{document}
 
\maketitle
\setcounter{page}{1}

\section{Introduction}\label{intro}

Heavy quarks can be produced by quark-antiquark interaction forming a virtual gluon decaying into a $c\bar{c}$ or $b\bar{b}$ pair. But the most important production mechanism is gluon fusion. Therefore, their spectrum is sensitive to initial parton distribution. The heavy quark can bind to a lighter quark and form open charm (D) or bottom (B). Due to the heavy mass of the reaction products, a large momentum is transfered during the process, corresponding to a short distance interaction leading to a small coupling constant $\alpha_{s}$ in the QCD framework. Their cross section can be calculated in QCD in terms of $\alpha_{s}^{2+n}$ (pQCD). {\it Fixed Order plus Next-to-Leading Logarithms} calculations (FONLL) \cite{Cacciari:2001td}, with $c \rightarrow D$ and $b \rightarrow B$ non-perturbative fragmentation functions, have shown reasonable agreement with Tevatron bottom \cite{Cacciari,Mangano} and charm \cite{Acosta} data at $\sqrt{s}=1.8$ TeV and midrapidity. Since heavy quarks are likely produced during early stages of a nucleus-nucleus collision and interact with the media, they can be used as a powerful probe for nuclear and deconfinement effects.\\

Nuclear modifications are generally accounted for by a constant $\alpha$ in $\sigma=\sigma_{N}A^{\alpha}$. There was no observation of any non-linear $A$ dependence ($\alpha \sim 1$) in open charm measurements at fixed target E769 (250 GeV $\pi$+A) \cite{E769}, WA82 (340 GeV $\pi$+A) \cite{WA82} and E789 (800 GeV $p$+A) \cite{Leitch} experiments. However, WA78 (320 GeV $\pi$+A in $x_{F}>0.2$, or small $x$ region) \cite{WA78} has shown a significant non-linear target dependence ($\alpha=0.81 \pm 0.05$) in D total cross section. Observations of charmonium suppression in $p$+A collisions \cite{E772,E866,NA50} and PHENIX d+Au data \cite{PHENIX} are relevant to the question of shadowing of the small $x$ gluon distribution in the nucleus \cite{Eskola}.\\

Some enhancement at high $p_T$ of hadron production relative to \pp collisions ($\alpha>1$) has been observed in SPS \cite{Aggarwal,Albrecht} and RHIC experiments \cite{Adler} some enhancement at hight . This $p_{T}$ broadening is known as Cronin effect \cite{Antreasyn} and it is attributed to the multiple scattering of the incident partons.\\

A strong high $p_T$ suppression has been observed in PHENIX Au+Au collisions at $\sqrt{s_{NN}}=$200 GeV at midrapidity for $\pi^{0}$ and charged particles \cite{Adler1,Adams1,Adler2}, despite the small enhancement in d+Au collisions measured in the same apparatus \cite{Adler,Adams2}. This behavior is well described by energy loss via gluon radiation in a hot and dense matter \cite{Mustafa,Lin}. The gluon radiation occurs for angles $\theta$ around the parton incidence higher than $\theta_c=m/E$. Therefore, heavy quarks should have less energy loss due to the dead cone around the incident parton \cite{Dokshitzer}.\\

Particle anisotropy is largely used to study the thermal conditions of the initial stage in non-central nucleus-nucleus collisions. In particular, the second component $v2$ of its Fourier expansion, called elliptic flow,  is sensitive to early pressure. Positive $v2$ is a indication of the particle thermalization \cite{Houvinen}. A heavy quark anisotropy measurement can test the mass dependence of the elliptic flow which is an indication of the media density.\\

The amount of charm remaining after the hadronization stage in A+A collisions is an important information to constrain the weight of open charm recombination contribution in $J/\psi$ spectra. Direct measurement of D mesons at RHIC in d+Au collisions with $p_T$ range up to 3 GeV/$c$ was reported by STAR Collaboration \cite{STAR}.\\

In this paper results from electron decays of heavy quarks obtained in \pp, d+Au and Au+Au collisions at $\sqrt{s_{NN}}$=200 GeV measured at mid-rapidity range in PHENIX Experiment will be shown. Most of the systematics is dedicated to extracting the photonic contribution from electron spectra by using two methods. Comparisons of the methods are presented for consistency considerations.\\

The data sets used in this paper can be seen at Table \ref{luminosity}.\\

\begin{table}[hb] 
\caption[]{Integrated luminosity used in each electron analysis at PHENIX.}\label{luminosity}
\begin{center}
\begin{tabular}{lllll}
\hline\\[-10pt]
Run & Projectile & Integrated\\
&& Luminosity \\ 
\hline\\[-10pt]
Run3 & \pp & 350 $nb^{-1}$\\
Run3 & d+Au & 2.74 $nb^{-1}$\\
Run4 & Au+Au & 241 $\mu b^{-1}$\\
\hline 
\end{tabular}
\end{center}
\end{table}

\section{Electron Measurement and Extraction of Heavy Quark Contribution.}
\subsection{PHENIX Apparatus and Electron Identification.}

\begin{figure}[htb]
   \centering
   \vspace{-.5cm}
   \insertplot{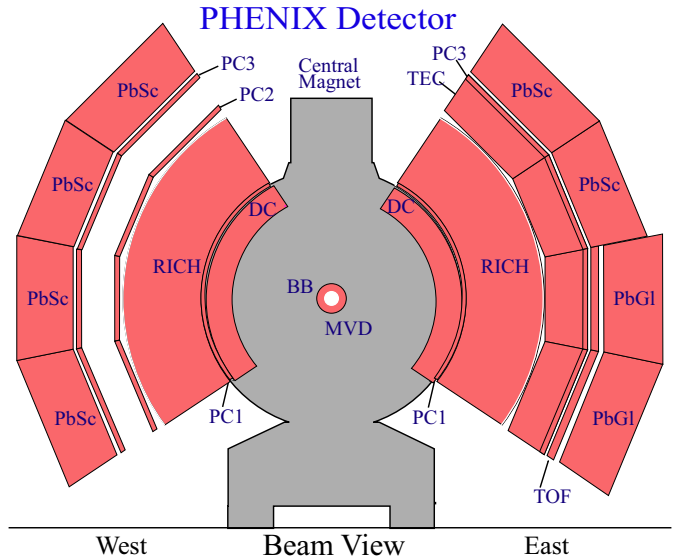}
   \vspace{-0.5cm}
   \caption[]{PHENIX central arm apparatus used for electron identification.}
   \label{PHENIX central}
\end{figure}

The PHENIX detector has two central arms spectrometers (Fig.\ref{PHENIX central}) covering a pseudo-rapidity $|\eta|<0.35$ and azimuthal angle $\Delta\phi=2 \times \pi/2$. The total geometrical acceptance is 35\%. A very important advantage in PHENIX central arms is the small amount of material. The total radiation length is $X_0=1.1\%$ with Multiplicity Vertex Detector (MVD), during Run1-3, and $X_0=0.4\%$ without MVD, from Run4 until now.\\

The collision is triggered by a Beam Beam Counter (BBC, $3.1 < |\eta| < 3.9$) and Zero Degree Calorimeter (ZDC, $|\eta| > 4$) coincidence in $92^{+2.5}_{-3.0}\%$ of Au+Au inelastic cross section. The multiplicity measured by the BBC and energy from neutrons detected in the ZDC provide the centrality measurement of the collision. Tracks for charged particles with momentum higher than 0.2 GeV/$c$ are recognized by the Drift Chamber (DC) and the pad chamber (PC1) with resolution $\Delta p / p \sim 1\%$. Only tracks matching the Electromagnetic Calorimeter (EmCal) clusters within 2 $\sigma_{pos}$ are used.\\

For the electron identification, we require a Cherenkov ring pattern around the track projection  on the Ring Cherenkov Detector (RICH) and an energy deposited in EmCal cluster $E$ corresponding to the particle momentum $^{+3\sigma_E}_{-2\sigma_E}$, since hadrons deposit only a fraction of their energy in the calorimeter. The remaining contamination from particles with momentum lower than the RICH pion threshold (4.9 GeV/$c$) comes from hadrons sharing Cherenkov rings with an electron. This contribution is statistically subtracted by using miscorrelated tracks (association of tracks with swapped Z coordinate in RICH).\\

The electron spectra are corrected taking into account geometrical acceptance and reconstruction efficiency. They are calculated in a detailed GEANT-based \cite{GEANT} simulation previously tuned with fully reconstructed beam pipe conversion electrons \cite{conversion} and the same dead areas as during data acquisition. The total electron acceptance times efficiency is almost uniform at 14\% for momentum above 1 GeV/$c$. The multiplicity effect in detection efficiency is estimated by embedding simulated electron raw signal into real data. The fraction of reconstructed simulated electrons corresponds to approximately 20\% for the most central events and 2\% for peripheral. These numbers are taken into account in the final efficiency.\\

The final inclusive electron spectra for \pp and Au+Au data is shown in figure \ref{cocktail}.\\

\subsection{Cocktail Method for Photonic Contribution Estimation.}

The relevant contributions for the electron spectra are :
\begin{itemize}
\item photonic source
	\begin{itemize}
	\item Dalitz decays of light neutral mesons or internal conversion : \\
	$\left(\pi^0,\eta^0,\omega^0,\eta'^0, \phi^0 \rightarrow 	\gamma e^+e^- \right)$
	\item conversion of photons from collision point in material (beam pipe, MVD, air) :
		\begin{itemize}
		\item light neutral mesons, i.e. : $\pi^0 \rightarrow \gamma\gamma \rightarrow \gamma e^+e^-$
		\item direct photons
		\end{itemize}
	\item virtual photons: $\gamma^* \rightarrow e^+e^-$
	\end{itemize}
\item weak kaon decay $K_{e3}~\left(K^{\pm} \rightarrow \pi^0e^{\pm}\nu_e \right)$
\item dielectron decays of vector mesons : $\rho, \omega, \phi \rightarrow e^+e^-$
\item thermal radiation
\item semileptonic decays of D, and B
\end{itemize}

\begin{figure}[htb]
   \vspace*{-0.5cm}
   \insertplot{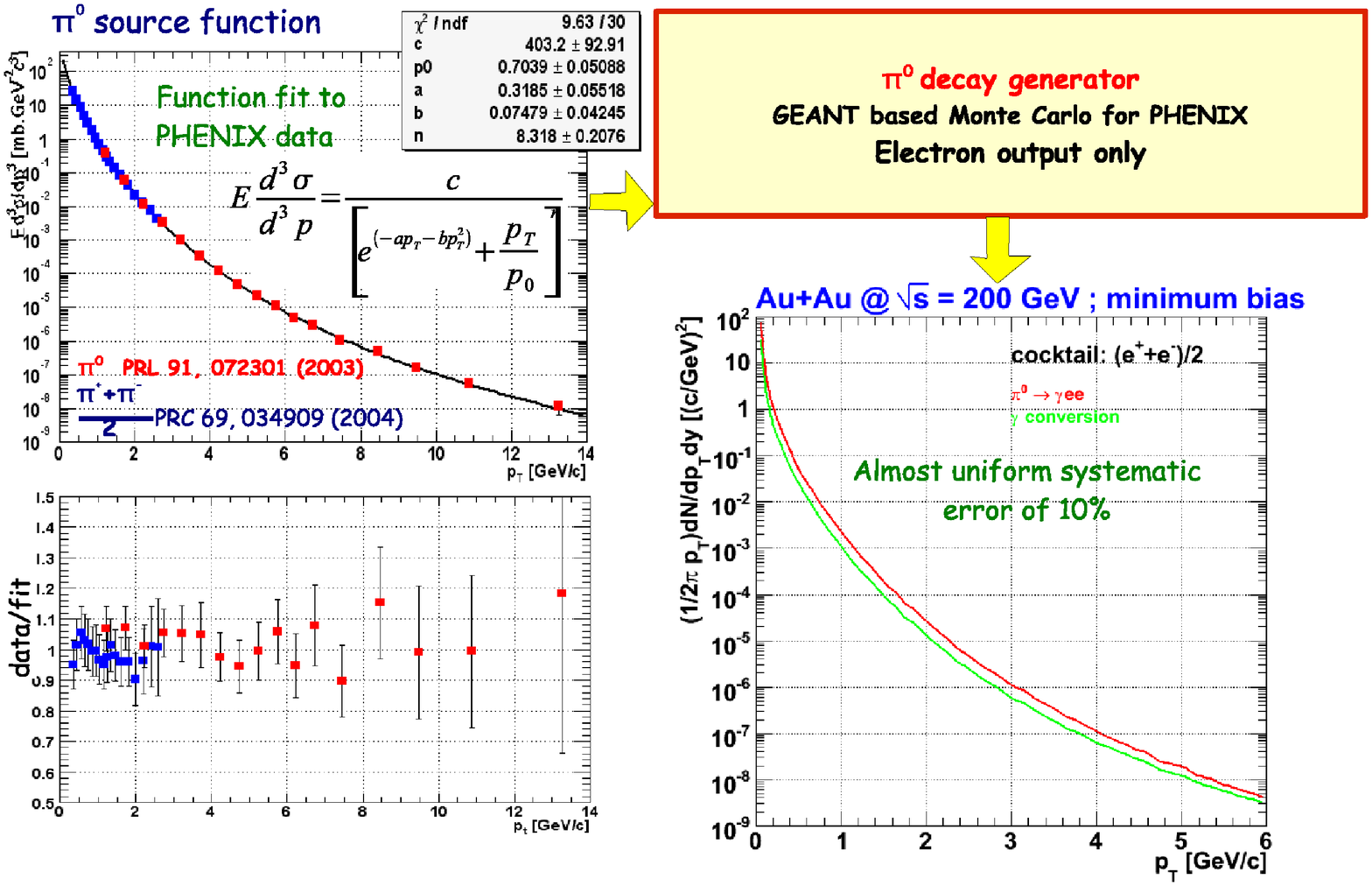}
   \vspace*{-0.5cm}
   \caption[]{Parameterization of $\pi^0$ spectrum function from $\pi^0$ \cite{Adler1} and charged hadron measurements \cite{hadron} at PHENIX.}
  \label{cocktail_pi0}
\end{figure}

The most important contribution comes from $\pi^0$. The function $E\frac{d^3\sigma_{\pi^0}}{d^3p}$ is parameterizated to $\pi^0$ and charged pion spectra measured at PHENIX (Fig. \ref{cocktail_pi0}). $E\frac{d^3\sigma_{\pi^0}}{d^3p}$ is an input for a GEANT based hadron decay generator including the PHENIX material for conversion estimation.\\

Assuming $m_T$ scaling and $h/\pi^0$ ratio constant at high $p_T$, we use $E\frac{d^3\sigma_{\pi^0}}{d^3p}$ for other light hadrons $h$ by replacing $p_T \rightarrow \sqrt{p_T^2+m^2_h-m^2_{\pi^0}}$ and normalizing the spectrum by the ratios :

\begin{itemize}
\centering
\item $\eta/\pi^0 = 0.45 \pm 0.10$
\item $\eta'/\pi^0 = 0.25 \pm 0.08$
\item $\omega/\pi^0 = 1.0 \pm 0.3$
\item $\rho/\pi^0 = 1.0 \pm 0.3$
\item $\phi/\pi^0 = 0.40 \pm 0.12$
\end{itemize}

\begin{figure}[htb]
   \centering
   \subfloat[\pp data \cite{singlepp}]
   {\hspace{-.5cm}\includegraphics[width=0.51\linewidth]{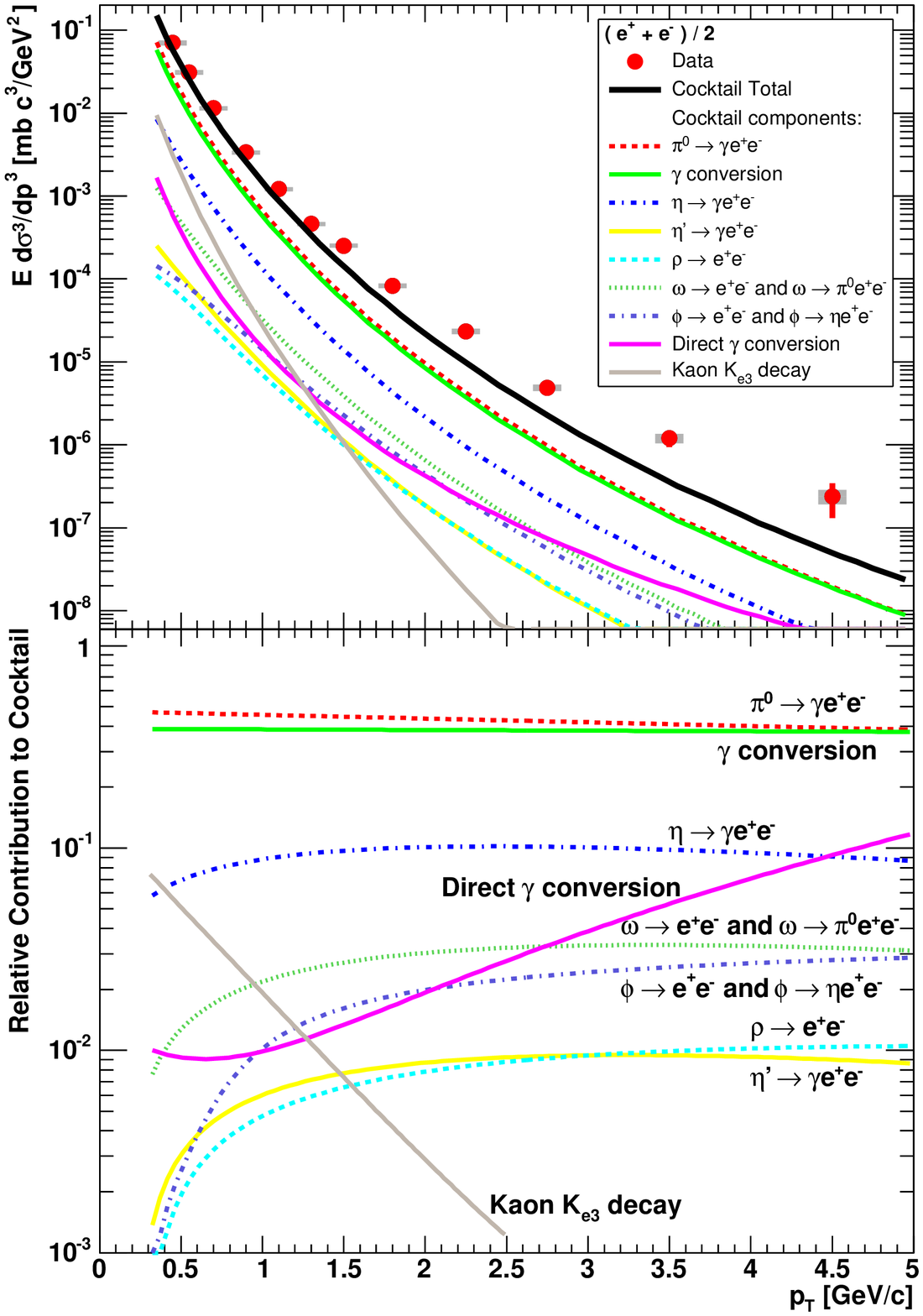}}
   \subfloat[Au+Au preliminary data]
   {\includegraphics[width=0.51\linewidth,height=0.5\textheight]{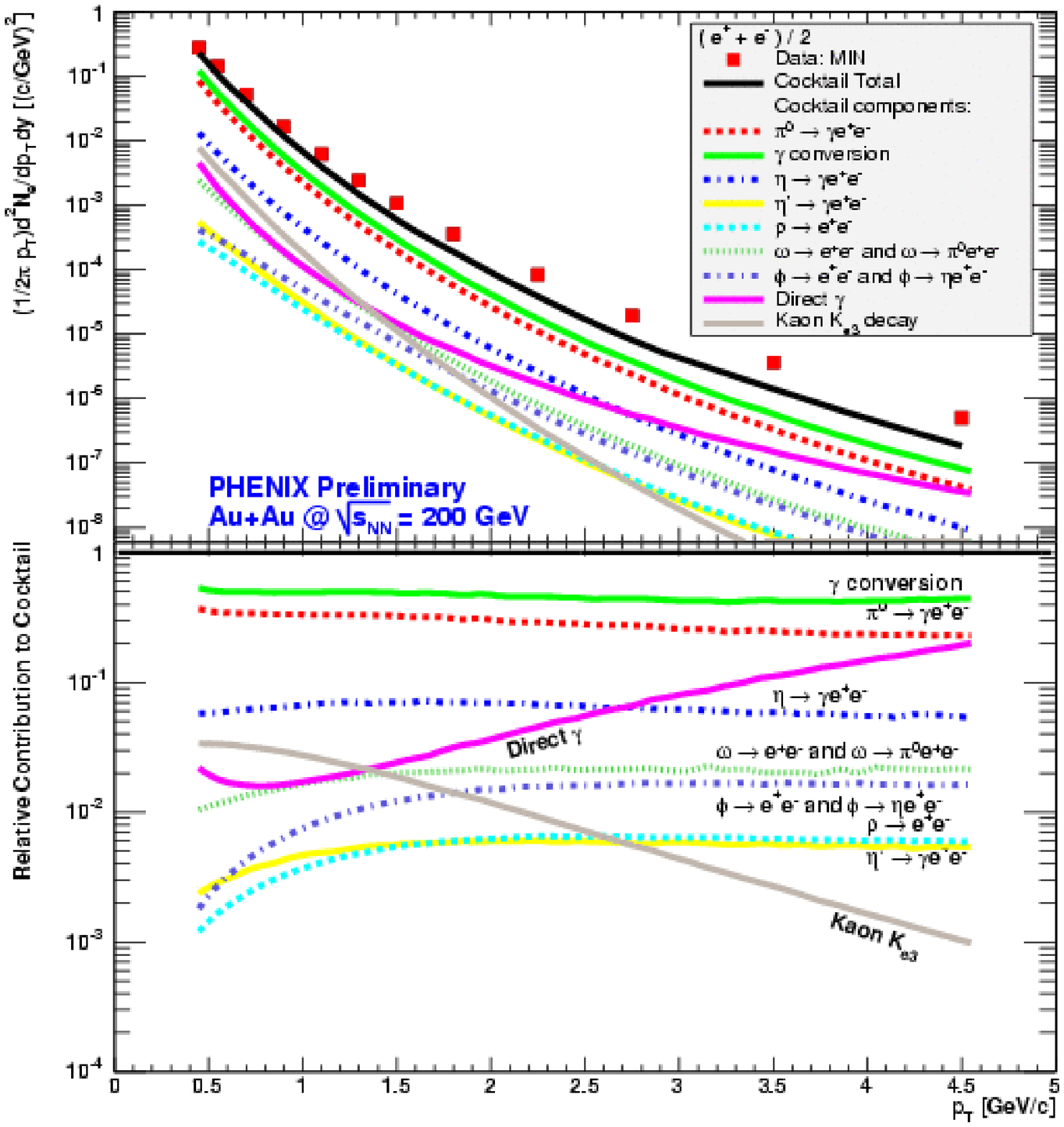}}
   \caption{Inclusive electron spectrum and relevant electron sources.}
   \label{cocktail}
\end{figure}

The spectrum obtained in this method agree whitin 20\% for all $p_T$ range in $\eta$ and $\phi$ yield measured at PHENIX . Direct photons \cite{direct photon} and kaon measurements \cite{hadron} at PHENIX are used in the $\gamma$ and kaon decay generator. All electron source cocktail result are shown in figure \ref{cocktail}.\\

\subsection{Converter Method.}
\label{converter}
The conversion method is based on the installation of additional material for a short period of each Run for photonic source estimation. The "electron converter" is a thin brass tube (1.7\% $X_0$) surrounding the beam pipe at 29cm.\\

The electron yields without ($N_e^{Conv-out}$) and with ($N_e^{Conv-in}$) the converter from photonic ($N^{\gamma}_e$) and non-photonic ($N^{non-\gamma}_e$) source are :

\begin{eqnarray}
N_e^{Conv-out} = ~~N^{\gamma}_e~~ + ~~N^{non-\gamma}_e~~~~~~~ \nonumber\\
N_e^{Conv-in}~ = R_{\gamma}N^{\gamma}_e + (1-\epsilon)N^{non-\gamma}_e \nonumber\\
R_{\gamma} = \frac{Y^{Conv-in}_e}{Y^{Conv-out}_e}
\end{eqnarray}

\begin{figure}[htb]
\centering
 \vspace{-0.5cm}
 \includegraphics[width=0.5\linewidth]{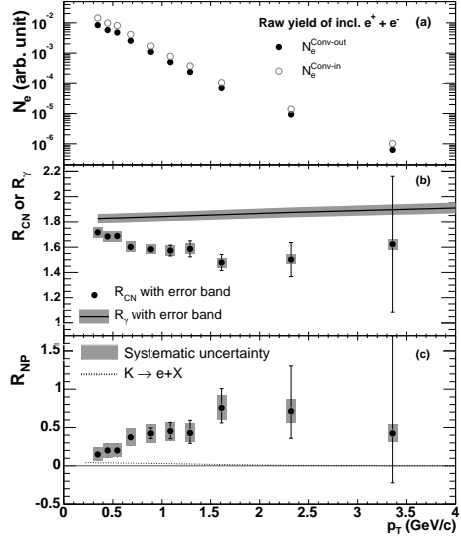}
  \caption{Data from Run2 \cite{singleAuAu} (a) $e^{\pm}$ yields with (open circles) and without (filled circles) converter. (b) Ratio of converter in/out $(R_{CN})$ $e^{\pm}$ yields and ratio of photonic yields $(R_{\gamma})$ determined by simulation. (c) Ratio of photonic/non-photonic $e^{\pm}$ yields $(R_{NP})$ and kaon contribution.}
 \label{conversion}
\end{figure}

During data acquisition with converter, $\epsilon\approx2.1\%$ of electrons are lost. The converter contribution factor $R_{\gamma}$ is estimated for each photonic source by a full GEANT based simulation with and without the converter. The simulation agrees whithin 2\% when compared to fully reconstructed conversion electrons in real data. The parameterizated function $E\frac{d^3\sigma_{\pi^0}}{d^3p}$ (Fig. \ref{cocktail_pi0}) is used for the $\pi^0$ input in this simulation. Likewise the cocktail method, $m_T$ scaling and $\eta/\pi^0$ ratio are applied to obtain the $\eta$ input. For the other mesons we assigned $R^h_{\gamma} = R^{\eta}_{\gamma}$, since their total contribution correspond to only 6\% of the photonic source at 3 GeV/$c$ and their Dalitz branching ratio are similar to $\eta$. The total $R_{\gamma}$ is a sum of the photonic sources weighted according to the hadron ratios.\\

\begin{figure}[htb]
\centering
   \includegraphics[width=1.0\linewidth]{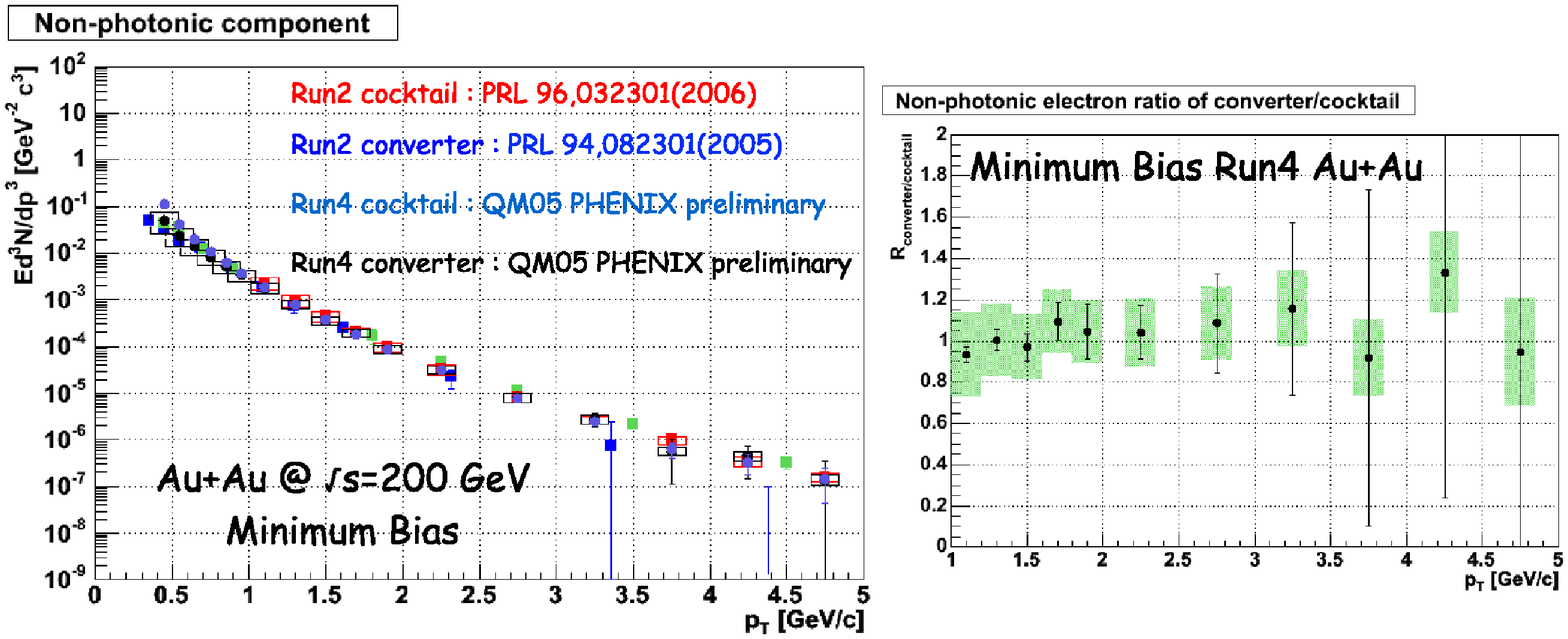}
   \caption{Photonic subtraction method and Run comparisons for Au+Au non-photonic electron spectra.}
   \label{comparison}
\end{figure}

Figure \ref{conversion} shows the momentum dependence of converter/non-converter yield $(R_{CN})$ and photonic/non-photonic yield $(R_{NP})$. The non-photonic yield contains contribution from weak kaon decay and vector mesons. These contributions are subtracted in the same manner as for the cocktail method. A good agreement is obtained between these different methods and data sets (Fig. \ref{comparison}).\\

\section{Heavy Quark Results.}

In this section only electron spectra and yields from semileptonic decays of heavy quarks are reported. These data are usually mentioned as non-photonic electron, since most of the subtracted electron contribution are internal and external $\gamma$ conversions. But in the data shown here, small contributions from kaons and vector mesons are also removed, according to the method described in the previous section.\\

\subsection{\pp  data and pQCD results.}

\begin{figure}[htb]
\centering
   \includegraphics[width=0.5\linewidth]{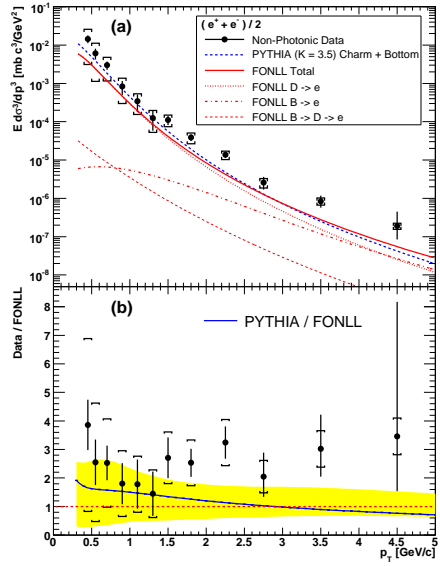}
   \vspace{-0.5cm}
   \caption{(a)Non-photonic electron spectrum measured in \pp collisions with PYTHIA and FONLL \cite{Cacciari2} predictions. (b) Data and FONLL ratio including error band and PYTHIA comparison.}
   \label{pp}
\end{figure}

The heavy quark production obtained in \pp collisions is reported in \cite{singlepp}. In figure \ref{pp} the spectrum is compared to leading order PYTHIA and FONLL pQCD calculations.\\

The phenomenological parton shower program PYTHIA 6.205 used a modified set of parameters from central and bottom hadroproduction \cite{Adcox} and CTEQ5L parton distribution functions \cite{Lai}. In order to accomodate NLO contributions, we use a scale factor K=3.5 in leading order contribution. The calculation describes well the data up to 1.5 GeV/$c$.\\

Above 2 GeV/$c$ the electron spectrum suggests contributions beyond FONLL pQCD calculation \cite{Cacciari2}. Higher order terms and/or jet fragmentation can play a role in higher momentum range.\\

The total cross section of charm quark-antiquark pair production was estimated by using the electron spectrum shape the PYTHIA extrapolation. The PYTHIA function including charm and bottom contribution was fitted to data with $p_T>0.6$ GeV/$c$. Extrapolating the charm contribution to $p_T=0$ GeV/$c$, the mid-rapidity charm cross section obtained was $d\sigma_{c\bar{c}}/dy = 0.20 \pm 0.03(stat) \pm 0.11(sys)$ mb. The rapidity integrated cross section correspond to $\sigma_{c\bar{c}} = 0.92 \pm 0.15(stat) \pm 0.54(sys)$ mb. The same estimation, performed by FONLL, gives $\sigma_{c\bar{c}} = 0.256^{+0.400}_{-0.146}$ mb. The systematics of these calculations is described in \cite{singlepp}.\\

\subsection{"Cold" Nuclear Effects in d+Au results.}

\begin{figure}[!htb]
   \centering
   \subfloat[PRELIMINARY $60\% < centrality < 88\%$]
   {\hspace{-0.4cm}\includegraphics[width=0.54\linewidth]{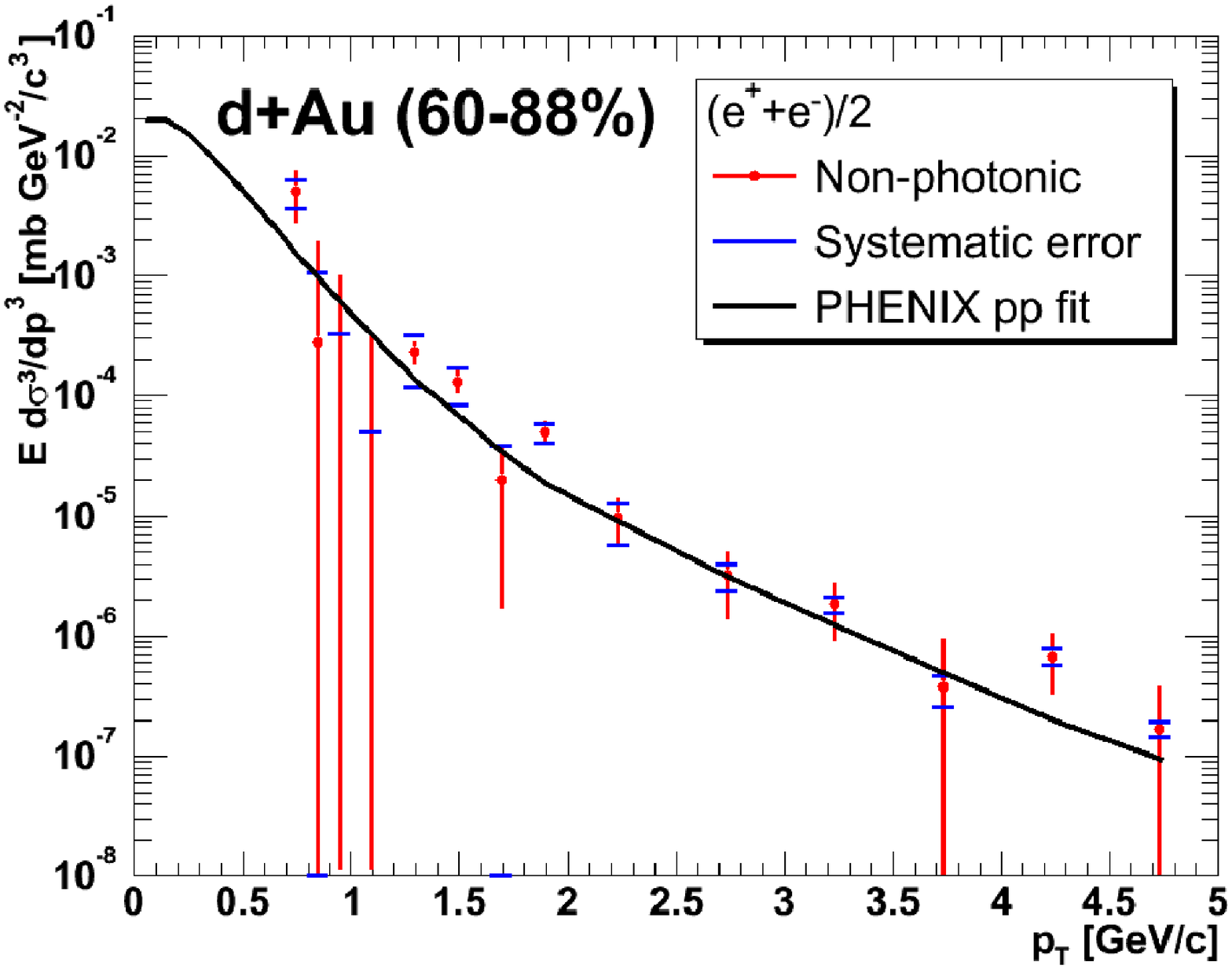}}
   \subfloat[PRELIMINARY $40\% < centrality < 60\%$]
   {\hspace{-0.8cm}\includegraphics[width=0.54\linewidth]{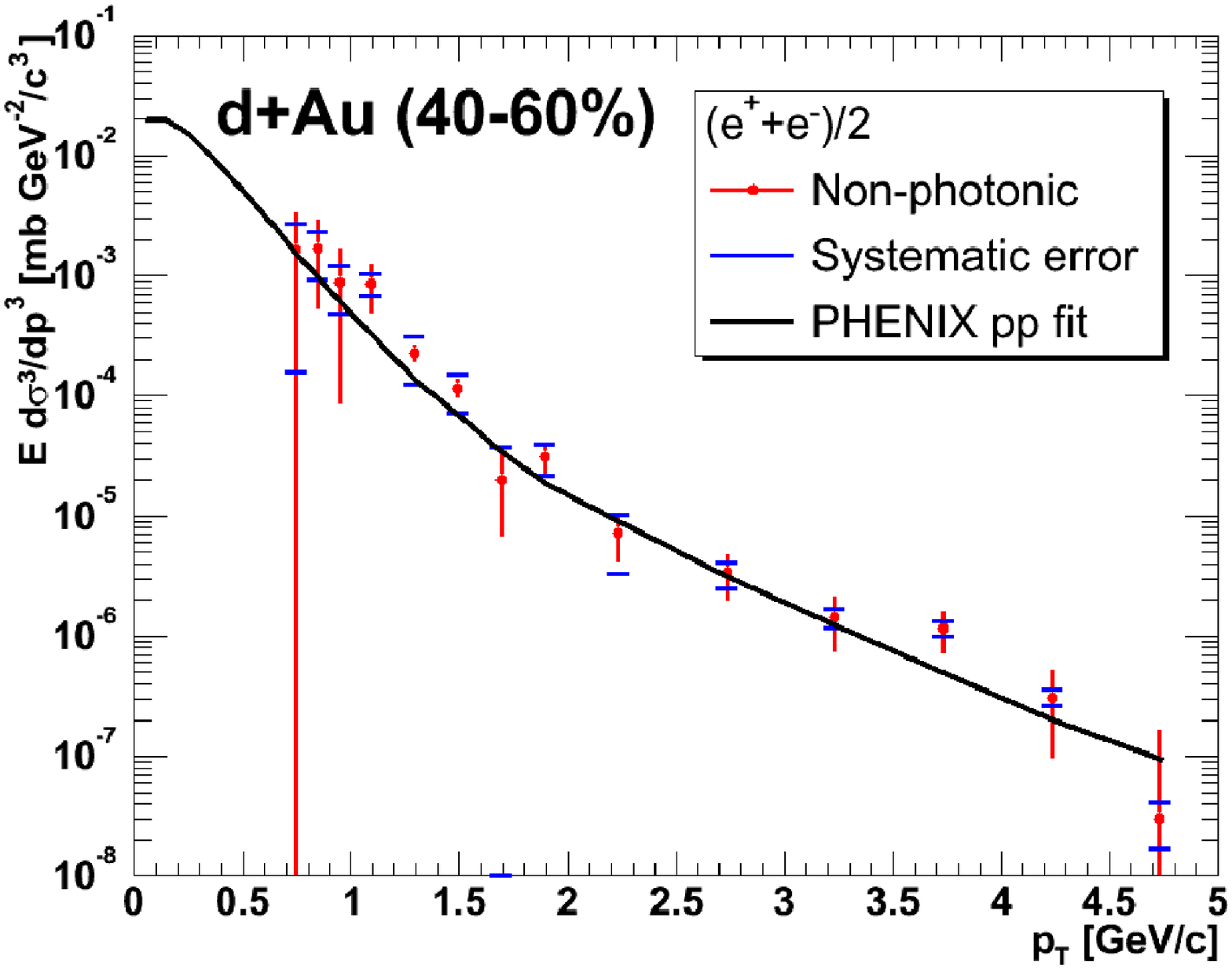}}
   \subfloat[PRELIMINARY $20\% < centrality < 40\%$]
   {\hspace{-0.4cm}\includegraphics[width=0.54\linewidth]{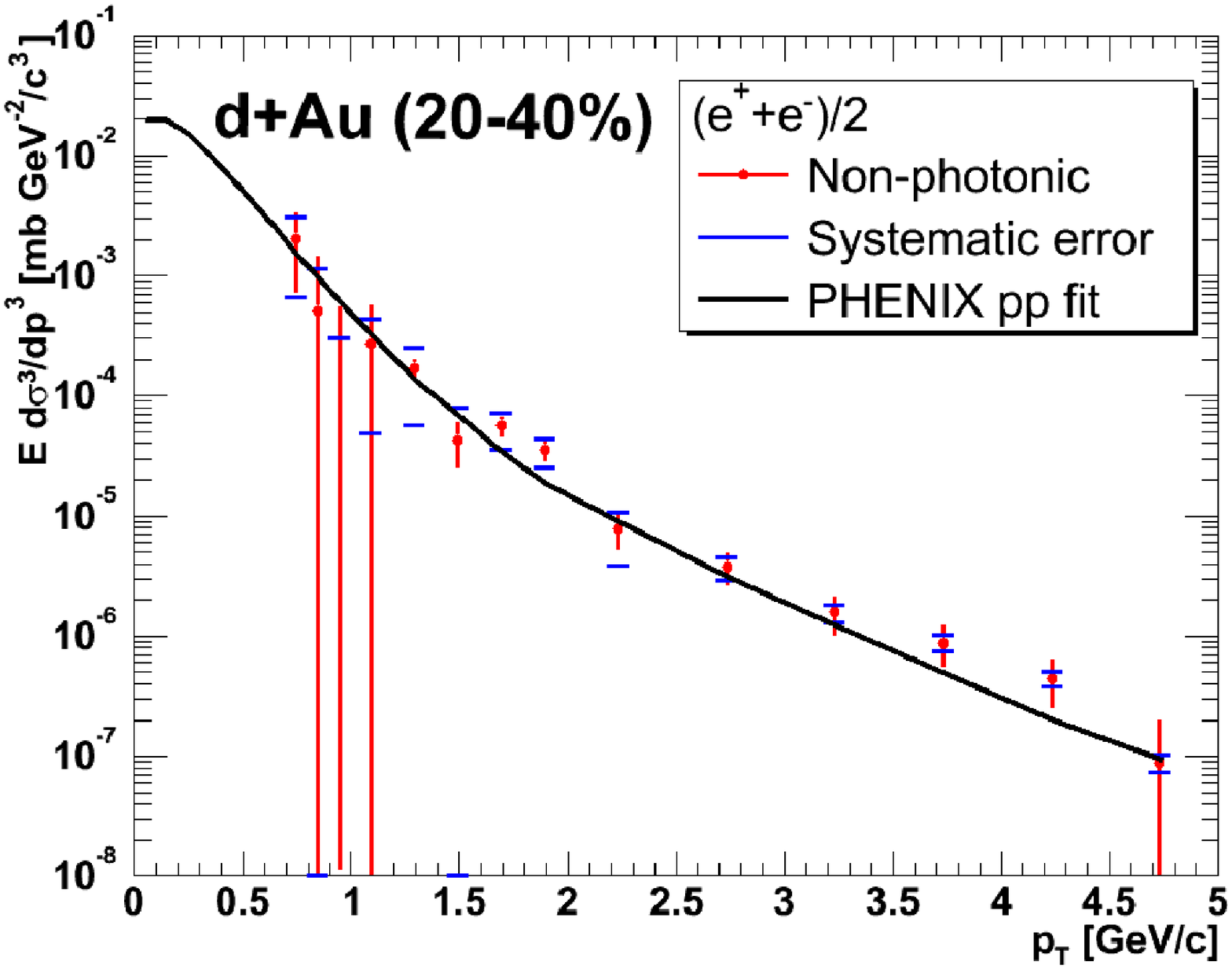}}
   \subfloat[PRELIMINARY $0\% < centrality < 20\%$]
   {\hspace{-0.8cm}\includegraphics[width=0.54\linewidth]{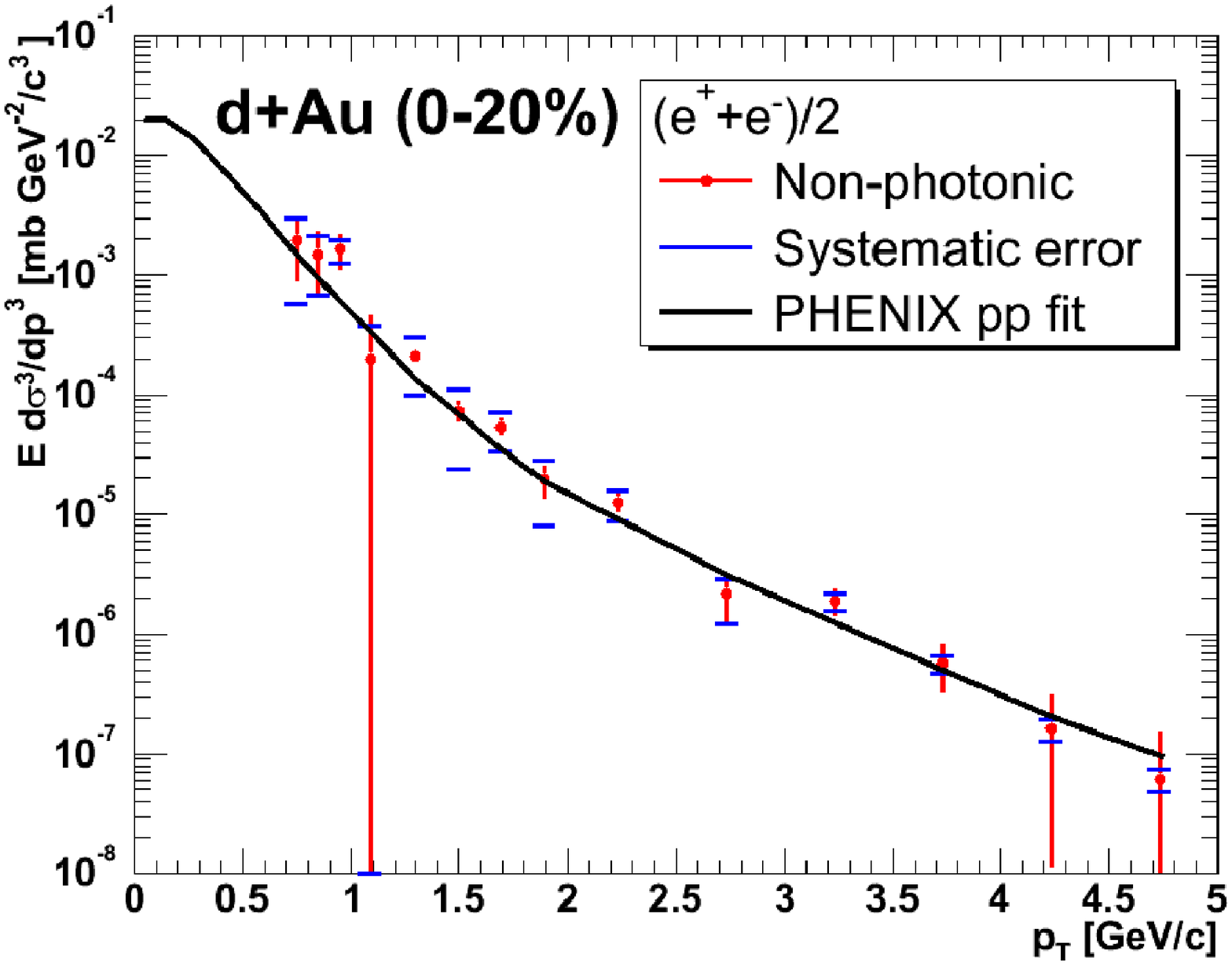}}
  \caption{Preliminary non-photonic electron invariant cross section per binary collision in d+Au for different centralities and \pp best fit \cite{QM4}.}
   \label{dAu}
\end{figure}

There is no indication of nuclear effect in d+Au collisions, since its preliminary non-photonic data scale with $N_{coll}$ (Fig. \ref{dAu}). The same behavior was observed in STAR fully identified D meson production \cite{STAR} and other experiments covering large $x$ region (see section \ref{intro}).

\subsection{Energy Loss in hot and dense matter formed in Au+Au collisions.}

\begin{figure}[!htb]
\centering
   \subfloat[]
   {\hspace{-0.4cm}\includegraphics[width=0.5\linewidth]{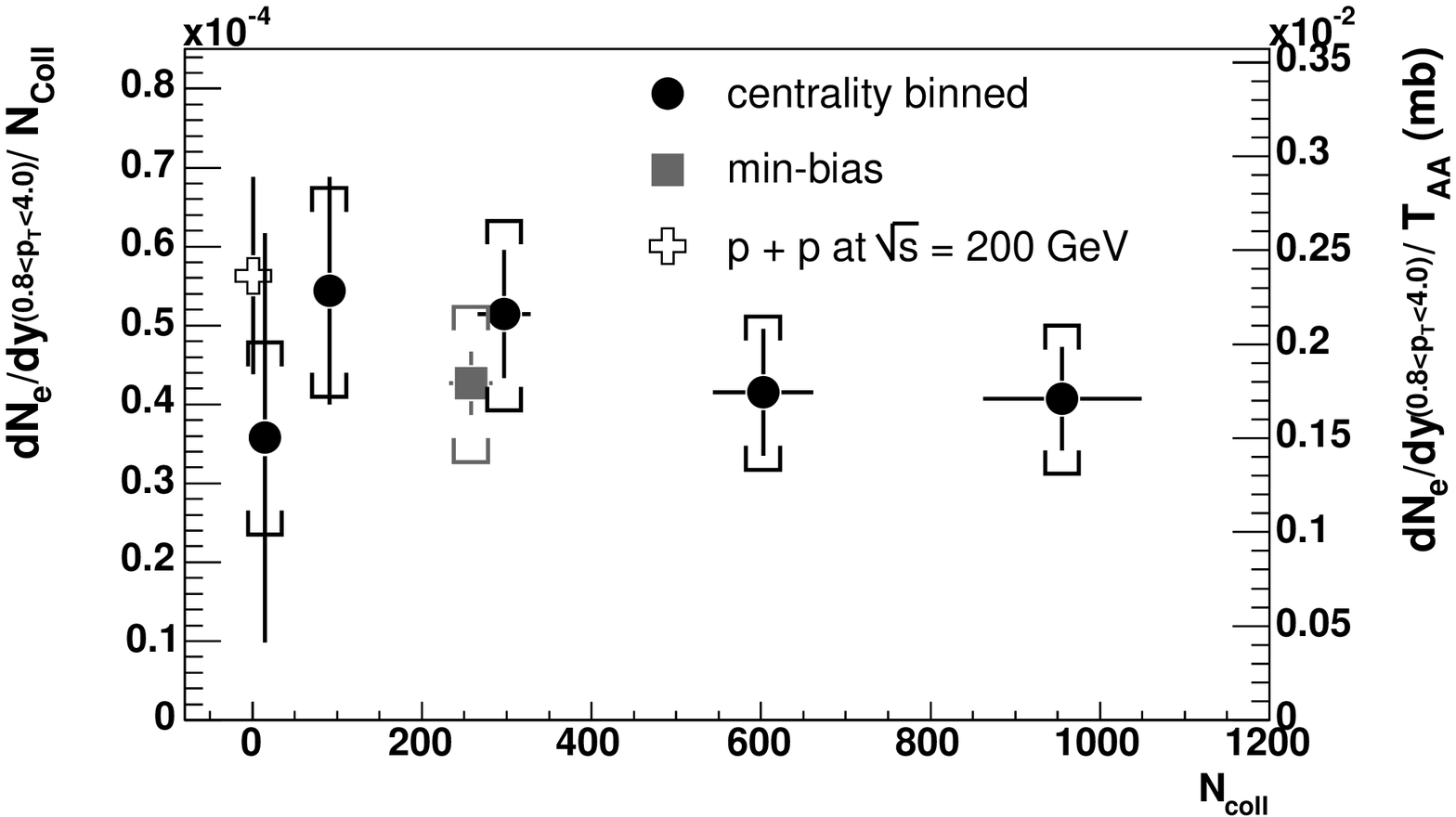}}
   \subfloat[]
   {\hspace{0.2cm}\includegraphics[width=0.5\linewidth]{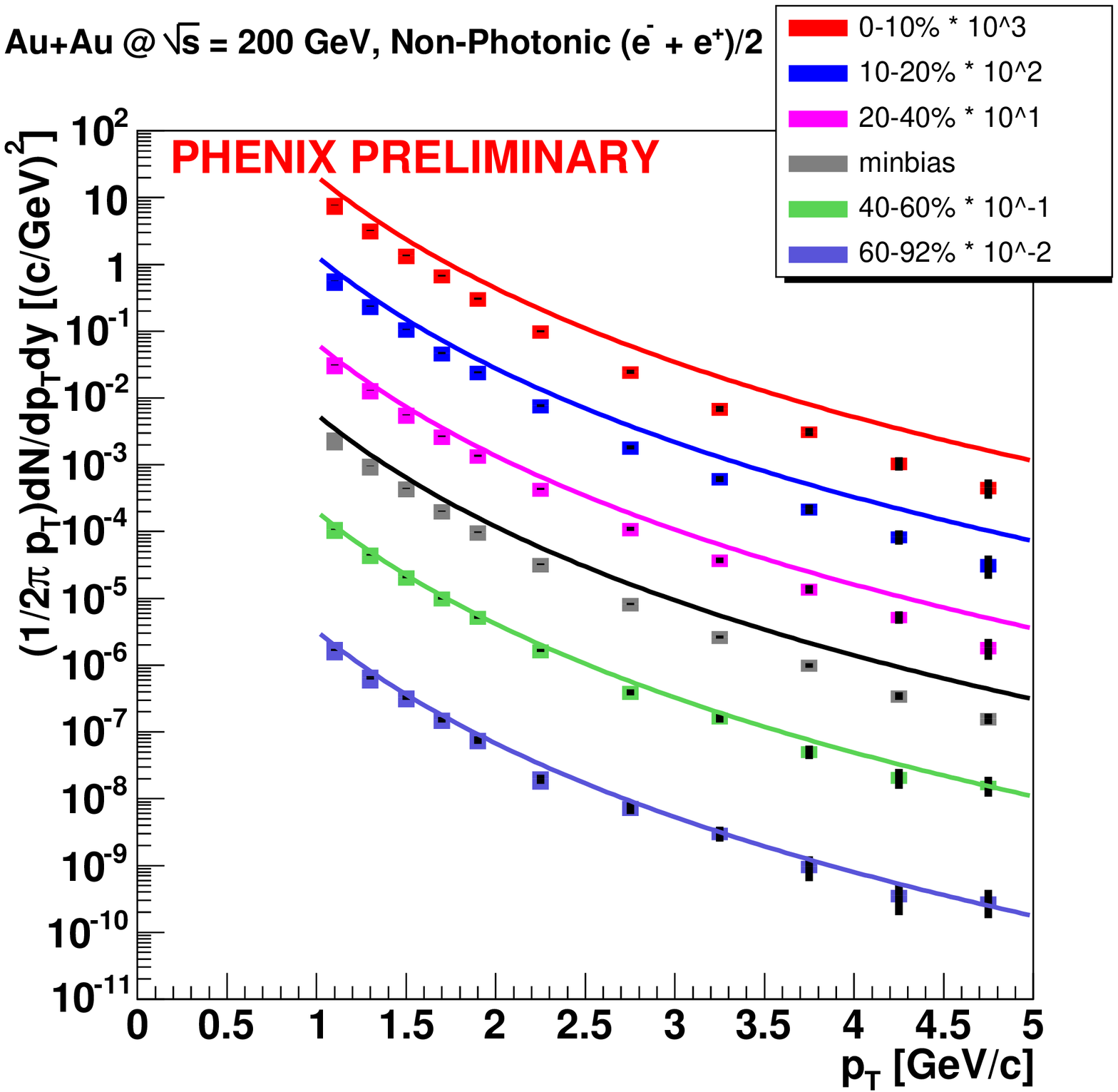}}
   \caption{Non-photonic electron : (a) Integrated yield (0.8 GeV/$c <$ momentum $<$ 4.0 GeV/$c$) {\it versus} $N_{coll}$ \cite{singleAuAu} \cite{singleAuAu}. (b) Run4 preliminary Au+Au spectra for different centralities compared to nuclear overlap integral $<T_{AA}>$ scaled \pp data.}
   \label{AuAu}
\end{figure}

The heavy quark yield at mid-rapidity range scales with $N_{coll}$ (Fig. \ref{AuAu}.a). This observation can be considered as an experimental verification of the binary scaling of a point-like pQCD \cite{singleAuAu}. However, medium effects can influence the momentum distribution.\\

\begin{figure}[!htb]
   \centering
   \subfloat[$60\% < centrality < 92\%$]
   {\includegraphics[width=0.33\linewidth]{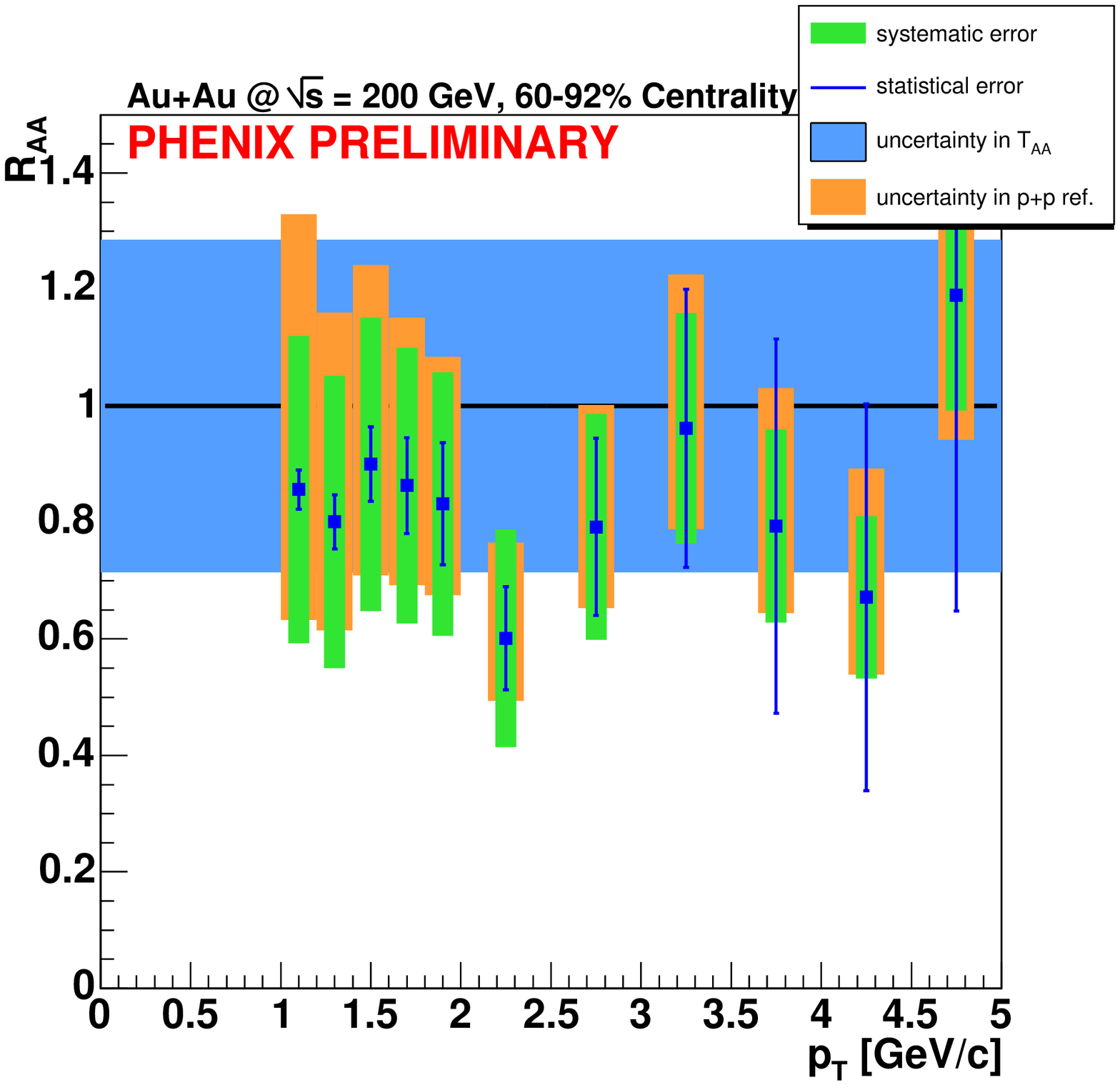}}
   \subfloat[$40\% < centrality < 60\%$]
   {\hspace{-0.2cm}\includegraphics[width=0.33\linewidth]{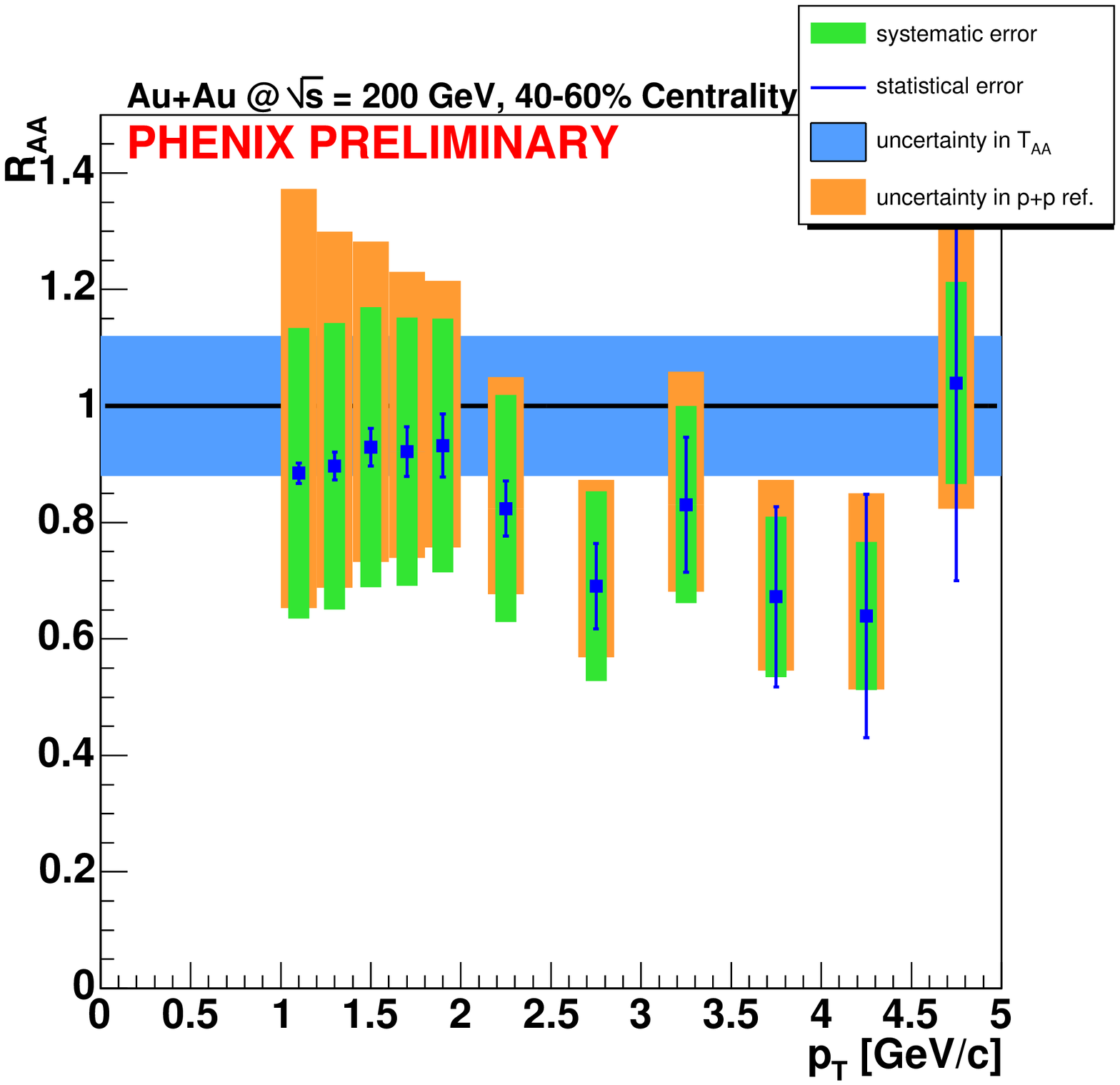}}
   \subfloat[$20\% < centrality < 40\%$]
   {\hspace{-0.2cm}\includegraphics[width=0.33\linewidth]{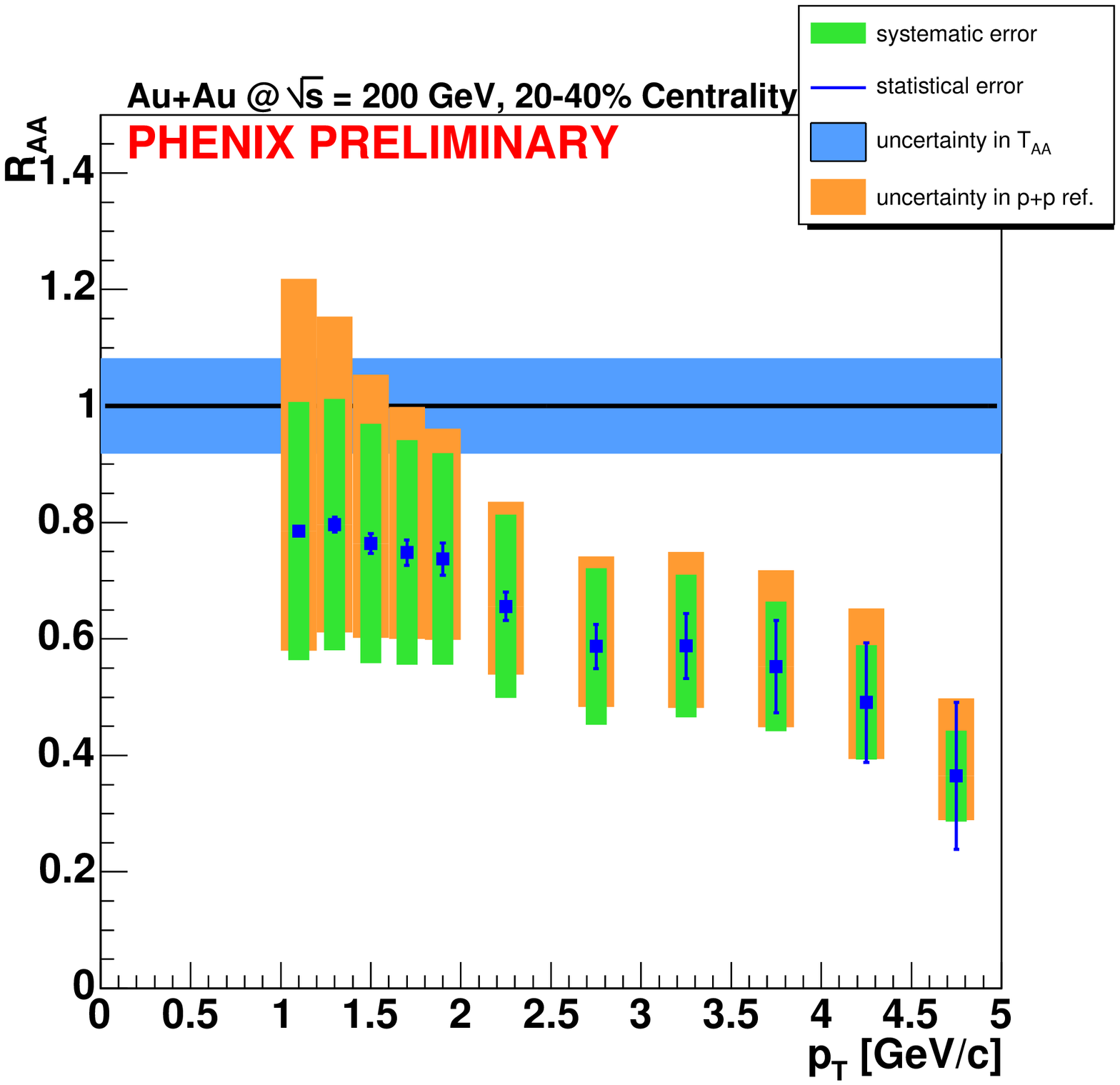}} \\
   \subfloat[$10\% < centrality < 20\%$]
   {\includegraphics[width=0.33\linewidth]{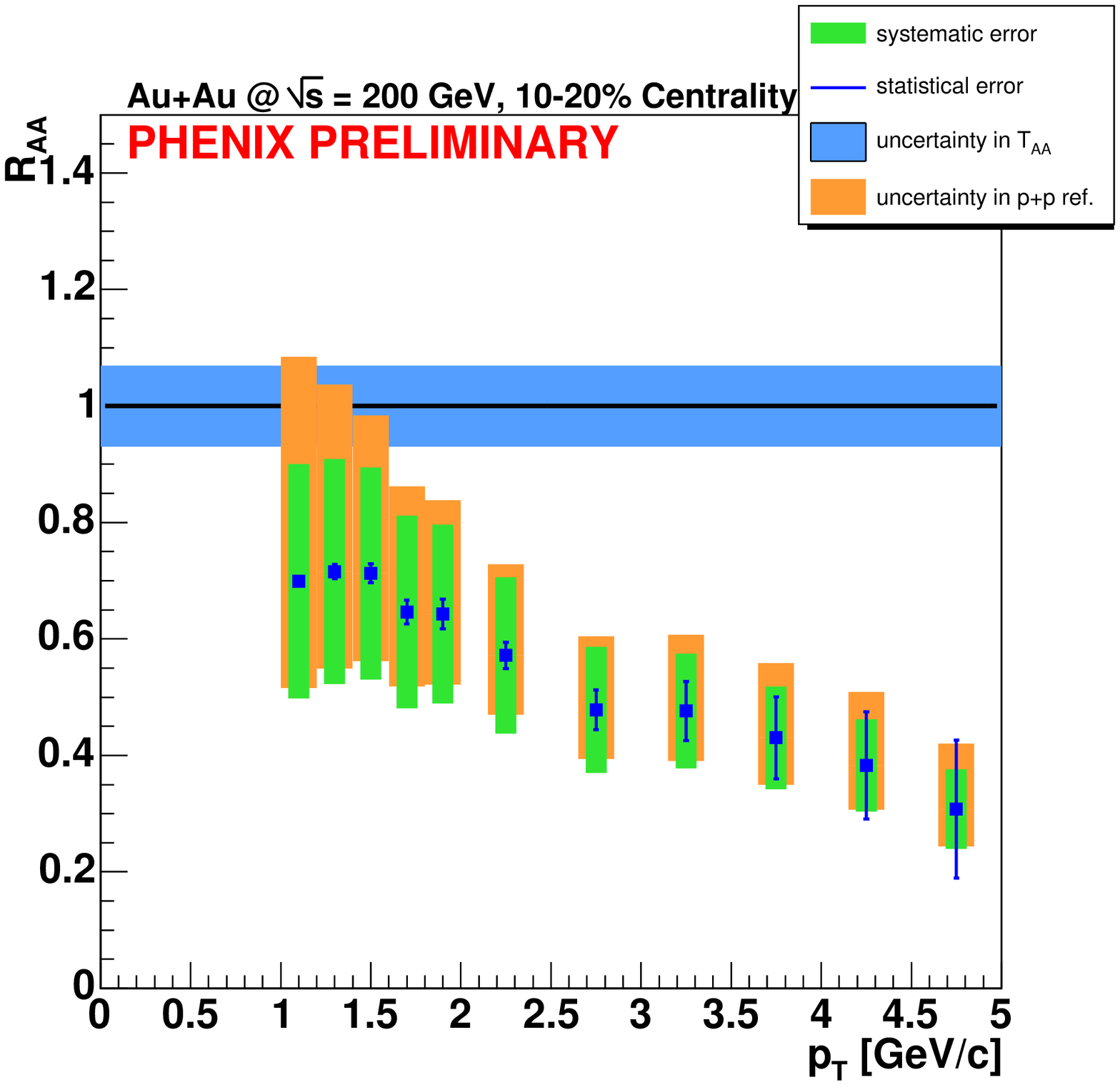}}
    \subfloat[\label{Raa_th}$0 < centrality < 10\%$]
   {\hspace{-0.2cm}\includegraphics[width=0.33\linewidth]{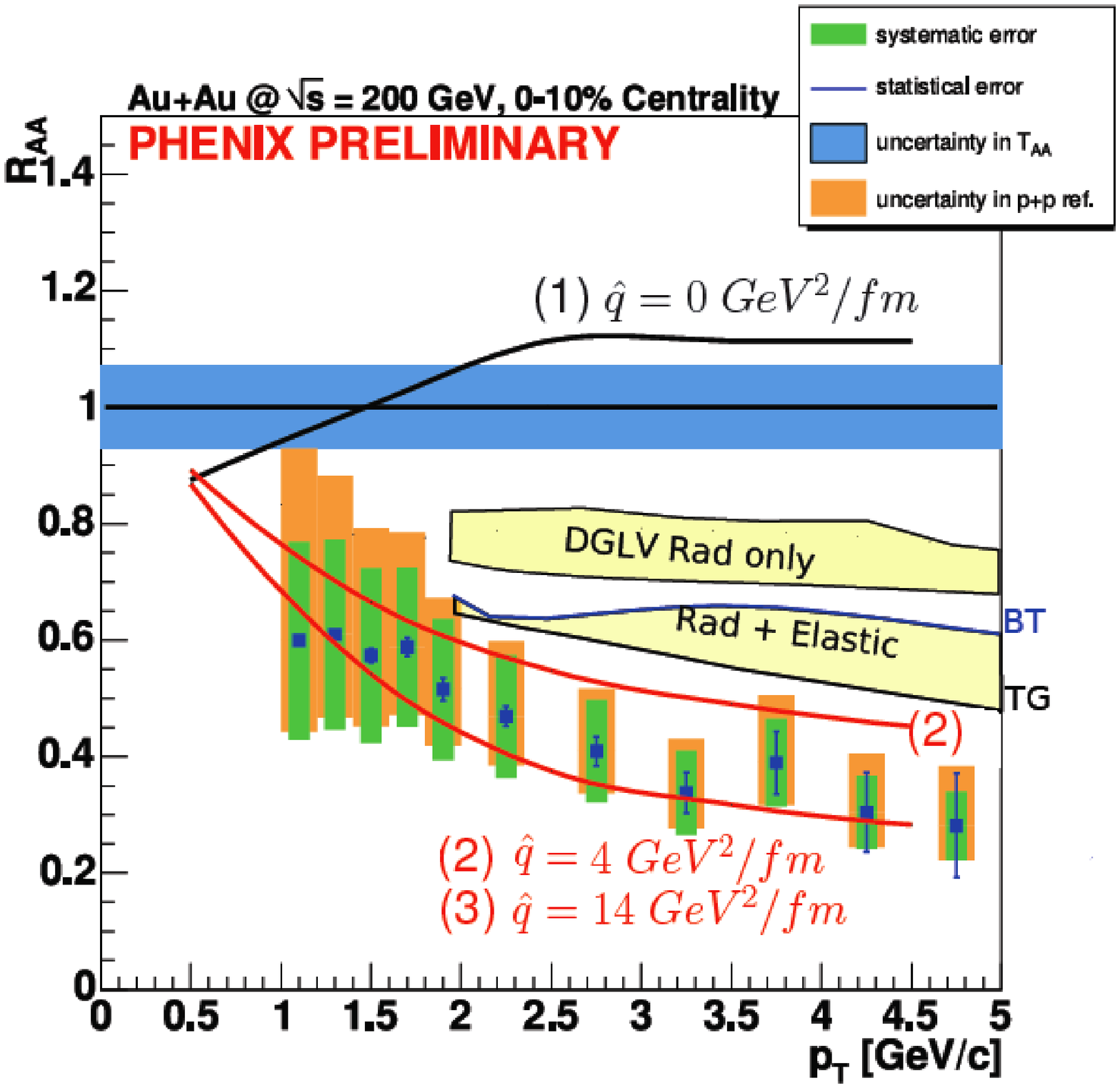}}
    \subfloat[Minimum bias]
   {\hspace{-0.2cm}\includegraphics[width=0.33\linewidth]{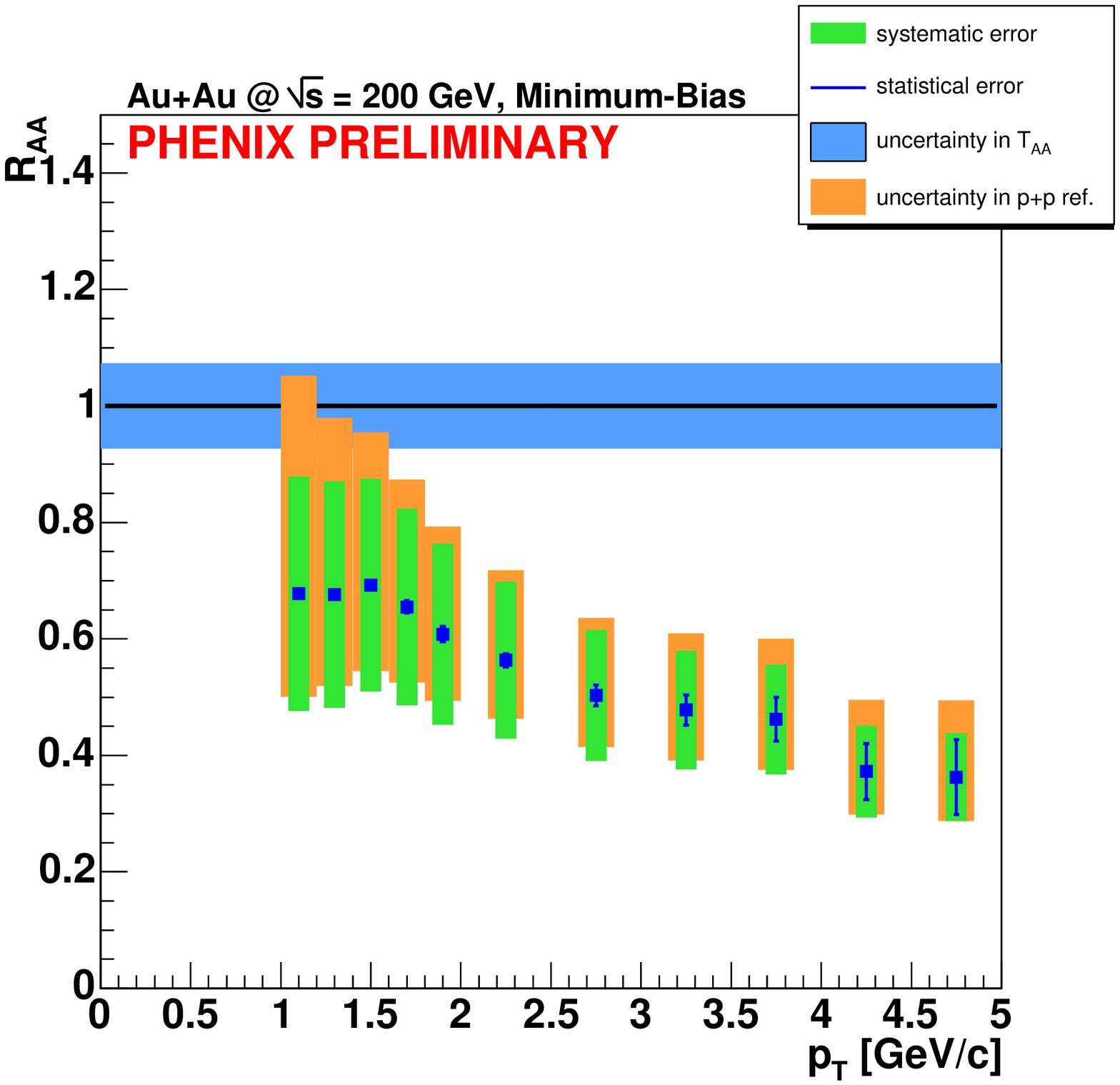}}
  \caption{Non-photonic electron $R_{AA}$ for different centralities. In panel (e) are shown theoretical expectations from charm (1-3) \cite{Armesto} and charm+bottom (radiation only and radiation+elastic bands) contributions \cite{Wicks}.}
   \label{Raa}
\end{figure}

Clear suppression is observed in central events at high $p_T$ when Au+Au data is compared to nuclear overlap integral $<T_{AA}>$ scaled \pp data (Fig. \ref{AuAu}.b). The quantification of this effect is obtained by $R_{AA} = \frac{dN_{Au+Au}}{\left<T_{AA}\right> \times d\sigma_{p+p}}$ calculations (Fig. \ref{Raa}).\\

For $p_T>$2 GeV/$c$ the heavy quark $R_{AA}$ is surprisingly similar to that for $\pi^0$ observed by PHENIX \cite{Adler1}, since the dead cone effect is neglected for lighter quark made mesons. It is commonly considered the time-average transport coefficient $\hat{q}$ denoting the average squared transverse momentum transferred from a hard parton per unit path length while traversing the medium. In the figure \ref{Raa_th} the spectrum in most central events and the charm energy loss expected for three different $\hat{q}$ \cite{Armesto} are presented.\\

Studies performed by S. Wicks {\it et al.} \cite{Wicks} including charm+bottom contribution and assuming gluon density $dN_g/dy = 1000$ are present in figure \ref{Raa_th}. In one of the scenarios the elastic energy loss is considered in the final state spectra besides the gluon radiation found in light meson jet quenching. The suppression does not seem to be affected by the expected significant contribution from bottom at momentum above 3 GeV/$c$.\\

\subsection{Elliptic Flow.}

\begin{figure}[!htb]
\centering
   \includegraphics[width=0.6\linewidth]{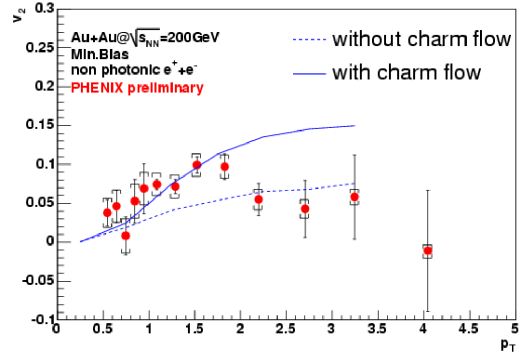}
   \caption{Non-photonic electron elliptic flow data and comparison with quark coalescence model with/without charm flow.}
   \label{v2}
\end{figure}

In electron anisotropy analysis the photonic contribution is estimated by the converter method (section \ref{converter}) for momentum below 1 GeV/$c$ and the cocktail method for momentum above 1 GeV/$c$, based on $v2$ from $\pi^0$ PHENIX data. The preliminary results and comparison with quark coalescence model \cite{Molnar} with/without charm flow are shown in figure \ref{v2}. This plot is consistent with published elliptic flow \cite{electronv2}, but with more statistics. In the charm $p_T$ range collective motion is observed as for light quark made mesons. The observed charm flow is consistent with partonic level thermalization and high density of the matter formed. The decreasing of $v2$ at high $p_T$ can be explained by a reduced flow from bottom dominance in this range.\\

\section{Outlook and Expectations.}

PHENIX detector measured heavy quark production for \pp, d+Au and Au+Au at mid-rapidity range for 0.8 $< p_T <$ 5 GeV/$c$ by using its electron decay channels. Photonic contribution to the electron spectra was calculated with two different methods which are in good agreement.\\

The next to leading order pQCD and PYTHIA calculations agree with \pp data at low $p_T$ and underestimates it at high $p_T$. The total charm-anticharm cross section is $\sigma_{c\bar{c}} = 0.92 \pm 0.15(stat) \pm 0.54(sys)$ mb and was estimated from a well parameterizated  PYTHIA extrapolation to \pp data.\\

The total non-photonic production scales with the number of binary collisions.  Medium effects were only observed in Au+Au collisions where the number of non-photonic electrons per number of collisions is suppressed at high $p_T$. The intensity of this suppression is comparable to what was observed for lighter mesons. This result represents a puzzle for the mass dependent gluon radiation scenario which explains reasonably well  the energy loss in $\pi^0$ and charged particle jet quenching observed at RHIC. The bottom contribution does not seem to affect the suppression.\\

Measurement of $v2$ points to a charm flow supporting a very high density in the early stages of the collisions. The reduced $v2$ at high momentum, despite the poor statistics in this range, can be explained by the contribution of a more massive particle, perhaps bottom.\\

Run4 RHIC luminosity allows us to reach a more extended $p_T$ region. Data from high luminosity Cu+Cu, \pp, and lower energy $\sqrt{s_{NN}}=$ 62.4 GeV collisions are coming soon. They will constrain production and medium effects models. The future installation of a vertex detector will provide a better signal/background non-photonic electron measurement and it will make possible to disentangle charm from bottom production, since B has a significant larger lifetime than open charm.\\

\section*{Note(s)} 
\begin{notes}
\item[a]
Permanent address: University of S\~ao Paulo, S\~ao Paulo, Brazil;\\ 
E-mail: slash@if.usp.br
\end{notes}

\vfill\eject

\begin{thebibliography}{9}
\bibitem{conversion}Photons from the collision point crossing the beam pipe can produce $e^+e^-$ pairs with orientation perpendicular to the magnetic field and form a peak in the invariant mass spectra. If one electron(positron) is carefully selected and it forms a pair with  a particle with the conditions above, the last particle is an unbiased positron(electron) which can be used for simulation tuning.
\end{thebibliography}
\end{document}